\documentstyle[12pt,bezier]{article}
\begin{document}
\title{\bf PHASE-LOCKING IN STRONGLY COUPLED SQUID CELLS}
\author{M. Basler\thanks{pmb@rz.uni-jena.de},
W. Krech\thanks{owk@rz.uni-jena.de} and
K. Yu. Platov\thanks{okp@rz.uni-jena.de} \thanks{
Permanent address:
Laboratory of Cryoelectronics,
Physics Department, Moscow State University,
Moscow 119899 Russia}\\
\sl Friedrich-Schiller-Universit"t Jena\\
\sl Institut fr Festk"rperphysik\\
\sl 07743 Jena\\
\sl Max-Wien-Platz 1}
\maketitle
\begin{abstract}
Within the RSJ model we performed
an analytical and numerical investigation of SQUID cells
consisting of two Josephson junctions
shunted by an extremely small
inductance leading to strong coupling of the elements.
Contrary to the well-known behavior of cells
shunted by a high inductance
voltage phases of the junctions are locked
with very small phase
difference for almost all values of external flux.
Only for external flux in the vicinity of half a flux quantum
phase difference rises rapidly to $\pi$.
\end{abstract}
\vskip-17cm
\hskip12cm
JENA-FKS-94-2
\vskip18cm

\centerline{Submitted to \sl Physics Letters A}
\vskip1cm

\hspace*{11cm}
Typeset with \LaTeX

\vfill

\eject
There has been growing interest in the development of complex
Josephson junction arrays during last years being caused by possible
applications in the field of satellite communication systems,
astronomical observations or construction of supercomputer chips [1-4].
Despite some promising
experimental results with 2D arrays [5-9]
detailed theoretical studies are still missing.
In addition, nearly all theoretical investigations
on coupled Josephson junctions in general are based on
weakly coupled elements. Otherwise, present day and near future
technology will allow construction of arrays with elements strongly
coupled by inductances around 3pH and smaller.

Thus before attacking more complex problems it is useful
to perform theoretical studies of very simple arrays of strongly coupled
Josephson junctions. One of the simplest arrays at all is
a completely superconducting SQUID cell consisting of
two identical Josephson junctions
with parallel current support inductively shunted by a very small
($L\ll\phi_0/2\pi I_C;\;\; \phi_0: \mbox{flux quantum},\;
I_C: \mbox{critical current}$) inductance.
Phase-locking in this model has been studied thoroughly
in the weak-coupling limit
several years ago \cite{Jain1,Likharev1,Krech1},
but any detailed theoretical investigations on
strong inductive coupling seem to be missing so far. The
results of our work show that the behavior of strongly coupled
Josephson junctions, especially under the influence of external
magnetic fields $\phi$, is qualitatively different from
what has been known for
weakly coupled elements.

The circuit to be considered here is shown in Fig. 1.

Both junctions, to be considered as being identical and
having negligible capacitance, are described within the RSJ model. The
behavior for weak coupling ($L\gg \phi_0/2\pi I_C$)
can be summarized as follows: (i) In any case there exists a stable
phase-locking regime. (ii) Without an external magnetic field the
difference between Josephson phases on both junctions vanishes in the mean.
In this case there is no
stationary current flow through the shunt. (iii) Any external flux $\phi$
leads to a non-vanishing difference of Josephson phases of the elements
causing a phase difference $\delta$ between voltages.
The mean value of this voltage phase difference $\delta$ can
be determined by the so-called reduced equation [1,10]
\begin{equation}
 \delta-\frac{1}{i_0\left(i_0+\sqrt{i_0^2-1}\right)}\sin\delta
 =\varphi,
\end{equation}

\begin{equation}
 i_0=I_0/I_C,\quad\varphi=2\pi\phi/\phi_0.
\end{equation}
Thus, for usual operation regimes with $i_0\approx1.5$ the
$\delta$-$\varphi$ dependence is nearly linear (cf. Fig. 2).

This may cause problems in
larger arrays; if parasitic magnetic fluxes cannot be suppressed
radiation output will be small because of cancellations
between contributions of different cells.

Here we consider the behavior for strong inductive coupling,
$L\ll \phi_0/2\pi I_C$. For vanishing McCumber parameter $\beta_C$
the circuit can be described by the RSJ equations
\begin{eqnarray}
 &&\dot{\phi_1}+\sin\phi_1=i-\l^{-1}(\phi_1-\phi_2+\varphi),\\
 &&\dot{\phi_2}+\sin\phi_2=i+\l^{-1}(\phi_1-\phi_2+\varphi)
\end{eqnarray}
where
\begin{eqnarray}
 &&\textstyle l=2\pi I_CL/\phi_0<<1
\end{eqnarray}
is the dimensionless inductive coupling and the dot denotes the
derivative with respect to the normalized time
\begin{equation}
s=\frac{2e}{\hbar}R_NI_Ct.
\end{equation}

For obtaining approximate solutions in the strong coupling limit
one introduces new variables
\begin{equation}
 \Delta=\phi_2-\phi_1\quad\mbox{and}\quad\Sigma=\phi_2+\phi_1
\end{equation}
and applies an expansion for small $l$
\begin{eqnarray}
  \Delta&=&\Delta_0+l\Delta_1+{\cal O}(l^2),\\
  \Sigma&=&\Sigma_0+l\Sigma_1+{\cal O}(l^2).
\end{eqnarray}
This analytical approach was complemented by a numerical investigation
exploiting the  Personal Supercomputer Circuit ANalyzer program PSCAN
\cite{Odintsov1,Polonsky1}
We will not go into the details of calculation here, but concentrate
on physical results.

For vanishing external flux the elements behave in the same way as
two weakly coupled elements; phase shift between voltages vanishes and
elements oscillate oppositely with voltages
\begin{eqnarray}
  v_1=v_2=\frac{\zeta_0^2}{i_0+\cos(\zeta_0s-\delta_0)},\\
  \zeta_0=\sqrt{i_0^2-1},\quad \delta_0=\mbox{const},
  \quad v_k=\frac{V_k}{I_CR_N}.
\end{eqnarray}
The behavior is quite plausible, because
in this case there is no current flow through the shunt.

If the external flux does not vanish, the behavior is more complicated.
To first order in $l$, time dependent voltage is obtained as follows:
\begin{eqnarray}\label{Loesung}
&&v_{1/2}=
\frac{\bar{\zeta}_0^2}{i_0+\cos(\varphi/2)\cos\bar{\zeta}_0s}
+l\frac{\bar{\zeta}_0\sin(\varphi/2)}{2(i_0+\cos(\varphi/2)
\cos\bar{\zeta}_0s)^2}\times\\\hfill\nonumber
\\
&&\Bigg[\sin(\varphi/2)\sin\bar{\zeta}_0s
\Bigg(i_0+\cos(\varphi/2)+
\frac{\bar{\zeta}_0^2}{\cos(\varphi/2)}
\ln\frac{i_0+\cos(\varphi/2)\cos\bar{\zeta}_0s}
{i_0+\cos(\varphi/2)}\Bigg)\nonumber\\
&&\mp\bar{\zeta}_0(\cos(\varphi/2)+i_0\cos\bar{\zeta}_0s)\Bigg]\nonumber
\end{eqnarray}
where
\begin{equation}
\bar{\zeta}_0=\sqrt{i_0^2-\cos^2(\varphi/2)}.
\end{equation}
This solution shows several specific features.
(i) As in the case of weak coupling there exists a solution exhibiting
phase locking for all values of the external field. (ii) For extremely
strong coupling $l\rightarrow 0$ both elements
behave similar to free junctions,
but with the current $i_0$
substituted by a kind of ''effective current''
depending on external flux,
$i_0/\cos(\varphi/2)$. (ii) There is a phase shift
between voltage phases being caused by the last term
of Eq. (\ref{Loesung}). However, contrary to the weak coupling
case, this phase shift is very small over a wide range of
external flux and only in the vicinity of the value
$\varphi=\pi$ the phase shift jumps to the value $\delta=\pi$.
Within the range of validity of our perturbation theory it is possible
to derive an explicit analytical expression for the phase shift $\delta$:
\begin{equation}\label{Phase}
\cos\delta=
\frac{(b^2+a_1a_2)}
{\sqrt{b^4+b^2(a_1^2+a_2^2)+a_1^2a_2^2}},
\end{equation}
where $a_1,\,a_2$ and $b$ are the leading Fourier coefficients of
the voltages (\ref{Loesung})
\begin{eqnarray}
a_{1/2}&=&-2\sqrt{\frac{i_0-\bar{\zeta}_0}{i_0+\bar{\zeta}_0}}\mp
l\sin(\varphi/2)\frac{\bar{\zeta}_0}{i_0+\bar{\zeta}_0},\\
b&=&l\sin^2(\varphi/2)\Bigg(\frac{i_0-\bar{\zeta}_0}{i_0\bar{\zeta}_0}
+\frac{1}{\bar{\zeta}_0}\sqrt{\frac{i_0-\bar{\zeta}_0}
{i_0+\bar{\zeta}_0}}
+\frac{\bar{\zeta}_0^2}{4i_0^2(i_0+\bar{\zeta}_0)}
\sqrt{\frac{i_0-\bar{\zeta}_0}{i_0+\bar{\zeta}_0}}\Bigg).
\end{eqnarray}

Fig. 3 shows a plot of phase shift between junctions against
external field.

This can be compared with results from  numerical simulation
(including a small shunt capacitance $\beta_C=0.01$; cf. Fig. 4).

The limiting case $\varphi=\pi$ is of special interest:
\begin{eqnarray}
  \textstyle v_1&=&i_0
  -\frac{l i_0}{2}\sin (i_0s-\delta_0)
  +\frac{l}{4}\sin2(i_0s-\delta_0),\\
  \textstyle v_2&=&i_0
  +\frac{l i_0}{2}\sin (i_0s-\delta_0)
  +\frac{l}{4}\sin2(i_0s-\delta_0).
\end{eqnarray}
This clearly indicates the phase difference between both voltages
to first order being equal to $\delta=\pi$.

Comparison of Figs. 3 and 4 shows that
even for $l\approx1$ where our analytical approximation is not longer
valid results are not far from that of numerical
simulation. Both figures show that already in this case the behavior is
qualitatively different from what has been known for weakly coupled
elements (dotted lines in Figs. 3 and 4).

Within the interval $0\le \varphi<\pi$ Eq. (\ref{Phase})
can be rewritten as
\begin{equation}
\delta(\varphi)=\arctan(a_2/b)-\arctan(a_1/b).
\end{equation}
This shows, that for $l\ll 1$, but finite, the solution
can be approximated by a step function
\begin{equation}
\delta\approx\pi\theta(\varphi-\varphi^*)
\end{equation}
with the step at
\begin{equation}
\varphi^*=\pi-i_0l
\end{equation}
(cf. Fig. 5).

Clearly $\varphi^*\rightarrow 0$ for $l\rightarrow0$. This approximation
might be useful considering more complicated arrays.

To conclude, our investigation of two strongly coupled Josephson
junctions has shown, that for every value of the external field
there exists a regime showing phase-locking between voltage phases.
Contrary to the case of weak coupling via large inductances,
phase shift between voltages is negligibly small
for a wide range of external flux for small
inductive shunts, but shows a narrow peak centered
around the value $\varphi=\pi$.
If larger arrays will show a similar behavior,
which has to be investigated in more detail, this would make
them nearly insensible to parasitic external flux.

\eject
{\bf Figure Captions}
\vskip1cm

\noindent
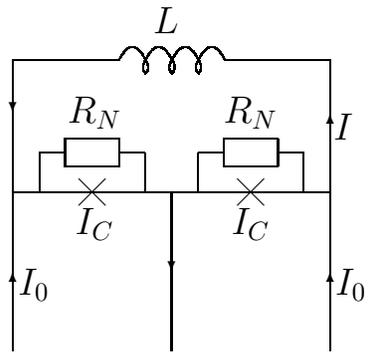
{\small Fig 1. Circuit of two Josephson
junctions with superconducting inductive SQUID coupling.}
\vskip1cm

\noindent
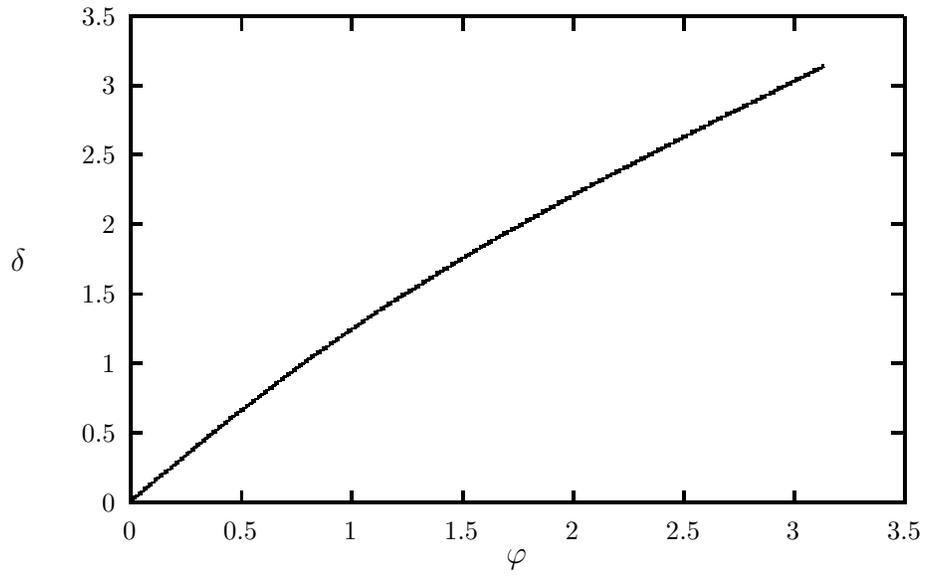
{\small Fig 2. Mean voltage phase
shift $\delta$ against
normalized external field $\varphi$ from analytical approximation
for weak inductive coupling ($i_0=1.5$).}
\vskip1cm

\noindent
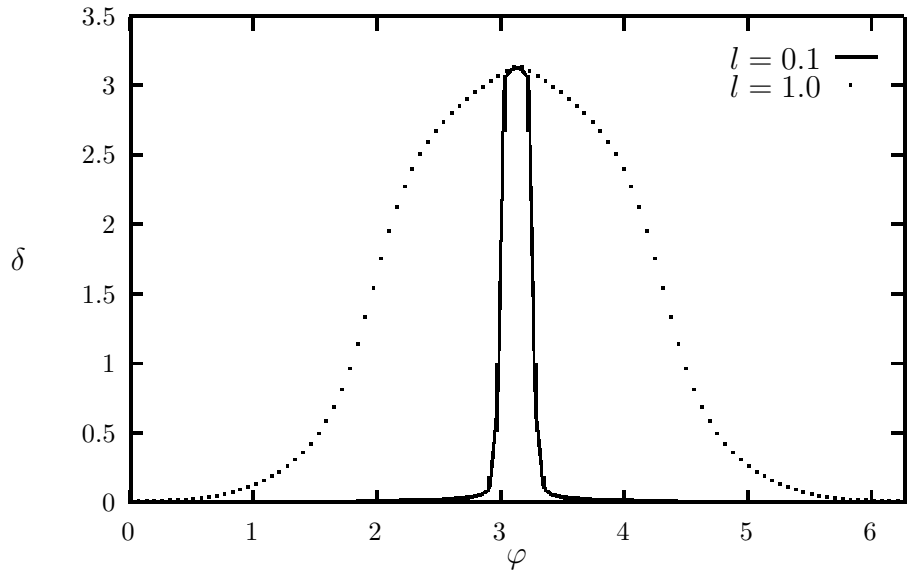
{\small Fig 3.
Mean phase shift $\delta$ against
normalized external filed $\varphi$ for strong inductive coupling
$l=0.1$ and medium inductive coupling $l=1.0$
obtained from analytical approximation (\ref{Phase})
($i_0=1.5$).}
\vskip1cm

\noindent
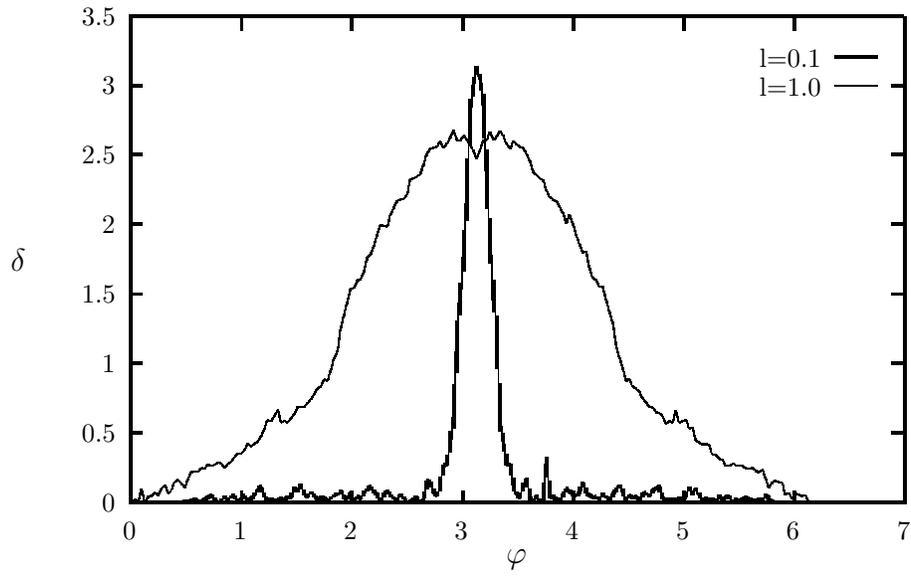
{\small Fig 4.
Mean phase shift $\delta$ against
normalized external filed $\varphi$ for strong inductive coupling
$l=0.1$ and medium inductive coupling $l=1.0$
obtained by numerical simulation ($i_0=1.5$).}
\vskip1cm

\noindent
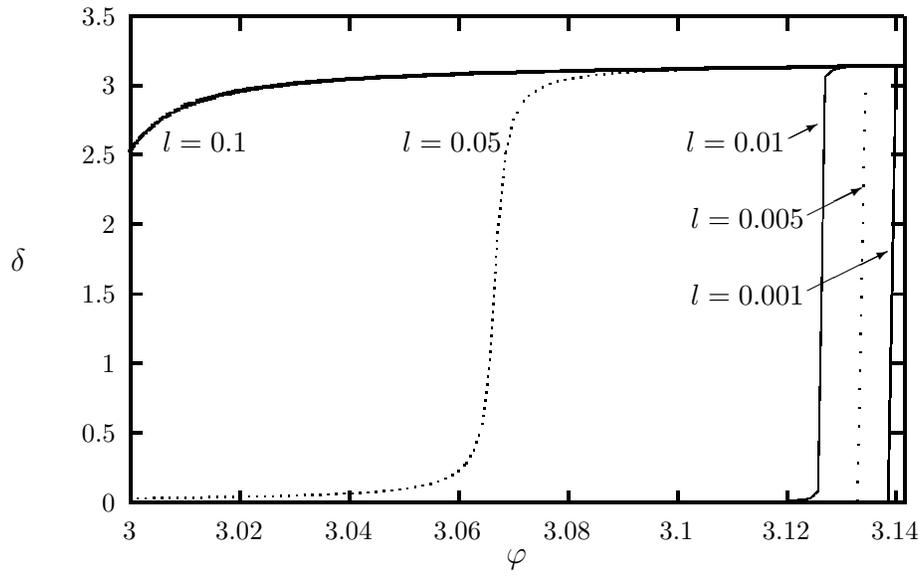
{\small Fig 5.
Mean phase shift $\delta$ against
normalized external field $\varphi$ for extremely
strong coupling ($0.001\le l\le 0.1$)
obtained from analytical approximation (\ref{Phase})
($i_0=1.5$).}

\eject
Fig.1
\vskip3cm
\centerline{

}
\end{document}